\documentclass[prb, twocolumn, superscriptaddress, showpacs,amsmath,amssymb]{revtex4}

\usepackage{float}
\usepackage{graphicx}
\usepackage{subfigure}

\begin{document}

\title{Millisecond spin-flip times of donor-bound electrons in GaAs}
\author{Kai-Mei C. Fu}
\email{kaimeifu@stanford.edu}
\author{Wenzheng Yeo}
\author{Susan Clark}
\affiliation{Quantum Information Project, SORST, JST, Edward L.
    Ginzton Laboratory, Stanford University, Stanford, California
    94305-4085, USA}
\author{Charles Santori}
\affiliation{Quantum Science Research, Hewlett-Packard
Laboratories, 1501 Page Mill Road, MS1123, Palo Alto, California
94304}
\author{Colin Stanley}
\author{M.C. Holland}
\affiliation{Department of Electronics and Electrical Engineering,
Oakfield Ave, University of Glasgow, Glasgow, G12 8LT, U.K.}
\author{Yoshihisa Yamamoto}
\altaffiliation[Also at ]{National Institute of Informatics,
Tokyo, Japan} \affiliation{Quantum Entanglement Project, ICORP,
JST, Edward L.
    Ginzton Laboratory, Stanford University, Stanford, California
    94305-4085, USA}

\begin{abstract}

We observe millisecond spin-flip relaxation times of donor-bound
electrons in high-purity $n$-GaAs.  This is three orders of
magnitude larger than previously reported lifetimes in $n$-GaAs.
Spin-flip times are measured as a function of magnetic field and
exhibit a strong power-law dependence for fields greater than 4~T.
This result is in qualitative agreement with previously reported
theory and measurements of electrons in quantum dots.

\end{abstract}

\maketitle

Since recent proposals for semiconductor spin-based quantum
information processing (QIP)~\cite{Kane98a, Loss98a, Imamoglu99a},
there has been a renewal of interest in the study of spin
processes in semiconductors.  The optical accessibility and the
possibilities of integrated device fabrication in III-V
semiconductors make electron spins in GaAs particularly promising
candidates for QIP. Theoretical research predicts long ($>$~ms)
spin-flip lifetimes~\cite{Woods02a, Khaetskii01a} and long ($\mu$s
- ms) decoherence times~\cite{Golovach04a, Semenov04a} for
electrons in GaAs quantum dots.  In this work we measure
millisecond spin-flip lifetimes (T$_1$) of donor electrons in
high-purity GaAs.  This long relaxation is comparable to the
longest reported value in quantum dots~\cite{Kroutvar04a} and is
three orders of magnitude longer than lifetimes reported in
$n$-GaAs samples with higher donor dopant
concentrations~\cite{Colton04b}.

Neutral donors (D$^0$) in semiconductors are in many ways natural
quantum dots.  At low doping concentrations $n \leq 10^{15}~
\textrm{cm}^{-3}$ and liquid helium temperatures, donor impurities
are non-interacting and bind a single electron.  This single
electron is always present in contrast to the quantum dot case
where the number of electron spins can be difficult to control.
Just as with quantum dots, it is possible to excite a bound
exciton, or electron-hole pair, at this site. The total bound
exciton (D$^0$X) complex is composed of two electrons in a
spin-singlet state, one hole, and the donor impurity.  Since each
donor electron is in the same environment and has the same
wavefunction (effective mass Bohr radius $a_B \approx
100~\textrm{\AA}$), there is little inhomgeneity and bulk optical
transition linewidths are as narrow as several GHz.  This is more
than three orders of magnitude narrower than quantum dots. The
homogeneity in D$^0$X systems makes them attractive candidates for
QIP applications where identical or nearly identical emitters are
necessary~\cite{Loock06a, Duan01a}. Extensive study has been made
of D$^0$-D$^0$X transitions~\cite{Karasyuk94a} and we used these
transitions to study T$_1$ of the bound electron.

%
%
\begin{figure}
\subfigure[] {\includegraphics[height =35mm,
keepaspectratio]{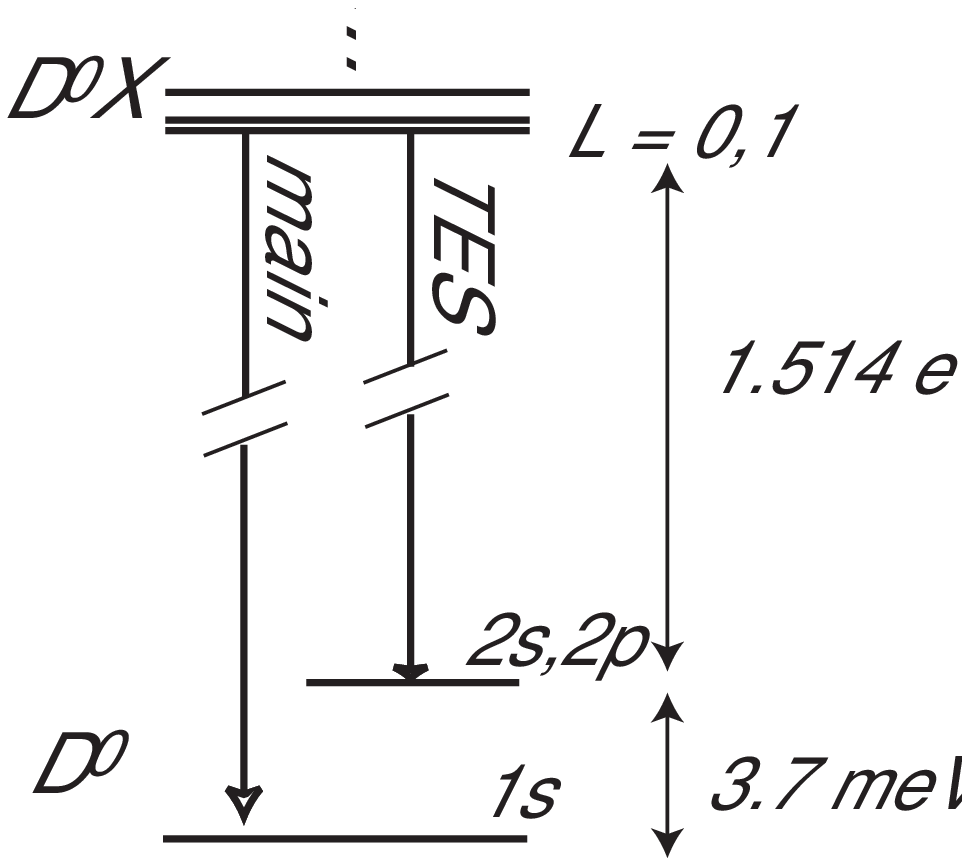}} \quad \subfigure[]{
\includegraphics[height = 40mm,
keepaspectratio]{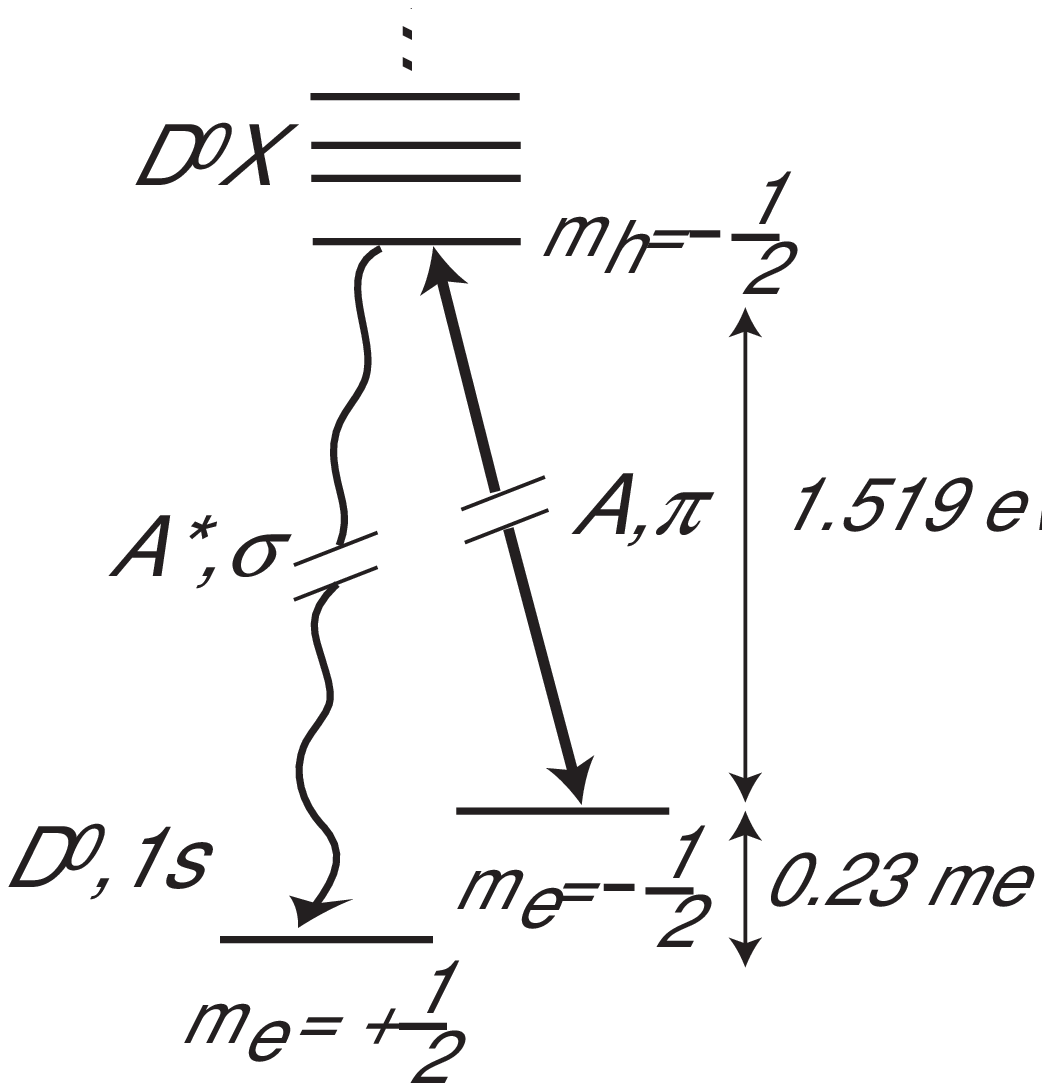}} \caption{\ (a) Energy level diagram
of the D$^0$X and D$^0$ states at 0~T.  In PLE experiments we
probe the excited state populations via the two-electron satellite
transitions. (b) Energy level diagram of the D$^0$ 1s state and
the lowest D$^0$X state in an applied magnetic field. To measure
T$_1$ we optically pump and probe the D$^0$ state using the $A$
($\pi$-polarized) transition. We detect the photoluminescence from
the $A^*$ ($\sigma$-polarized) transition. A complete energy
diagram of the GaAs D$^0$X states in a magnetic field can be found
in Ref.~\cite{Karasyuk94a}.}\label{energydiag}
\end{figure}
%
%

The sample consisted of a 10~$\mu$m GaAs layer on a 4~$\mu$m
Al$_{0.3}$Ga$_{0.7}$As layer grown by molecular-beam epitaxy on a
GaAs substrate. The sample had a donor concentration of
$\sim$5$\times$10$^{13}$~cm$^{-3}$. We mounted the sample
strain-free in a magnetic cryostat in the Voigt ($\vec{k} \perp
\vec{B}$) geometry.

\begin{figure}
{\includegraphics[height = 60 mm,
keepaspectratio]{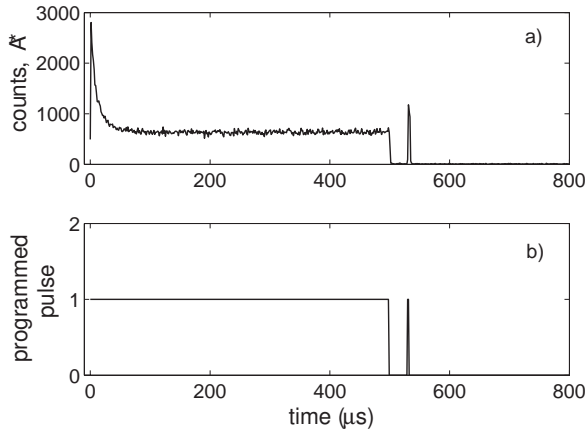}}\caption{(a) Typical data obtained
during the pump-probe experiment.  Time-resolved
photoluminensecence from the $A^*$ line is collected at 9.9 T.
Optical pumping can be observed during the first 20~$\mu$s of the
pump pulse.  Partial recovery is observed after the 30~$\mu$s
delay (b) Pump-probe programmed intensities.} \label{pumpprobe}
\end{figure}

An energy diagram of the D$^0$ and D$^0$X complexes in a magnetic
field are shown in Fig.~\ref{energydiag}b.  To measure T$_1$ of
the donor electrons  a long pulse is applied to the $A$
transition. The system is excited to the $|m_h =
-\frac{1}{2}\rangle$ bound exciton state where it then decays to
both electron spin states. After many cycles, this pumps electrons
from the $|-\frac{1}{2}\rangle$ state to the $|\frac{1}{2}\rangle$
state. After this pulse, the system evolves in the dark for a
variable time $\tau_\textrm{wait}$. During this time the electron
population returns to thermal equilibrium and the
$|-\frac{1}{2}\rangle$ state is repopulated with a characteristic
time T$_1$. A short pulse resonant on the $A$ transition then
probes the population in the $|-\frac{1}{2}\rangle$ state and the
intensity of the $A^*$ transition is measured.  The experiment is
then repeated with different time delays $\tau_\textrm{wait}$.  An
example pulse sequence and photoluminescence (PL) data are shown
in Fig.~\ref{pumpprobe}.

%
%
%
%

\begin{figure} [t]
{\includegraphics[height = 60 mm,
keepaspectratio]{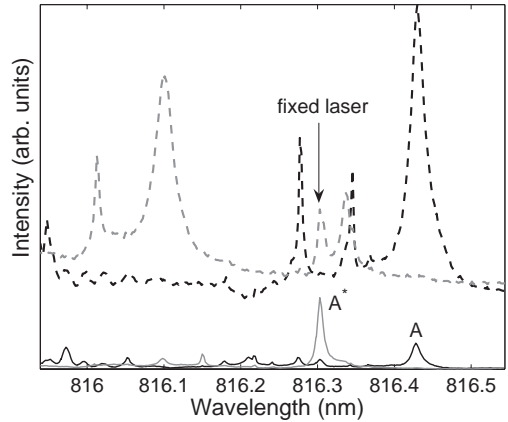}}\caption{Single laser PLE scans (solid
lines) and double laser PLE scans (dashed lines) at 9.9~T.  In the
single laser PLE scans, a $\pi$-polarized (black) and
$\sigma$-polarized (grey) laser is scanned over the D$^0$X
transitions. Photoluminescence (PL) is collected from an
$|m_h=-\frac{1}{2}\rangle$ TES transition and is proportional to
population in the D$^0$X $A$ state.  Overall intensity is weak due
to optical pumping.  In the two laser PLE scans, a second laser is
fixed resonantly on the $A^*$ transition as the first laser scans
over the D$^0$X transitions.  A large enhancement in PL intensity
is observed for transitions originating from the
$|-\frac{1}{2}\rangle$ state. } \label{PLE}
\end{figure}

In order to ensure that population is indeed being pumped into the
other electron Zeeman state and not to some state outside our
system, we perform single-laser and two-laser photoluminescence
excitation (PLE) scans described in previous work ~\cite{Fu05a}.
In the single laser PLE scan a continuous wave Ti:Sapphire laser
is scanned over the D$^0$X transitions as emission from the $|m_h
= -\frac{1}{2}\rangle$ state to its two-electron satellites (TES)
is detected (Fig.~\ref{energydiag}a). In this case as the laser
scans over each transition the photoluminescence is weak  due to
optical pumping. In a second scan a diode laser resonantly excites
the $A^*$ transition during the Ti:Sapphire scan.  The diode laser
repumps the trapped electrons back into the $|-\frac{1}{2}\rangle$
state allowing efficient population of the
$|m_h=-\frac{1}{2}\rangle$ state (Fig.~\ref{PLE}). An eight-fold
increase in the PLE intensity of the $A$-transition is observed.
This result indicates that electrons are being pumped between the
two Zeeman states and not to a state outside the system.

In the T$_1$ pump-probe measurement we resonantly excited the $A$
transition for a time $\tau_1 >$ 100 \textrm{$\mu$}s with a
Ti:Sapphire laser.  The intensities used in this experiment needed
to be extremely low (1-10 mW/cm$^2$) to minimize the dynamic
polarization of the nuclear spins within the donor electron
wavefunction~\cite{Fleisher84a}.  If larger powers were used a
diagmagnetic shift in the energy of the A line was observed of up
to 10~GHz, shifting the line off of resonance with the excitation
laser.  It is possible to estimate the nuclear field using a
phenomenological relation describing the diamagnetic shift of the
GaAs D$^0$X in a magnetic field \cite{Karasyuk94a}. At 9~T this
shift corresponds to a modest nuclear field of $B_{nuc} = 0.06$~T
and 1\% nuclear polarization.

During the pump down phase we monitored the excited state
population by spectrally filtering the $A^*$-transition.
Additionally, we reduced the collection of the scattered pump
light by collecting the opposite polarization.  We observed the
time-resolved decay in the PL from the $A^*$ transition as
electrons were pumped into the $|+\frac{1}{2}\rangle$ state
(Fig.~\ref{pumpprobe}).  The polarization of the $A$
($\pi$-polarized) and $A^*$ ($\sigma$-polarized) transitions
indicate that the resonant D$^0$X state corresponds to a hole-spin
of $m_h = -\frac{1}{2}$.  This assignment is consistent with
previous spectroscopic work with crystal axis $\langle 110\rangle$
parallel to the magnetic field~\cite{Karasyuk94a}.   In the
spherical approximation the corresponding branching ratio is 1:2
to the electron Zeeman states
$|+\frac{1}{2}\rangle$:$|-\frac{1}{2}\rangle$.  Although spherical
symmetry is reduced to tetrahedral symmetry by the crystal, we
still observe strong PL from both of these transitions
(Fig.~\ref{PLE}). With the fast (ns) radiative relaxation of the
D$^0$X state and the long ($>$100~\textrm{$\mu$}s) measured T$_1$
times, the intensity ratio of the optically pumped to the
recovered populations should be close to 0.  However the observed
PL ratio never exceeded 1/8. This ratio increases at lower field
and indicates that the effective T$_1$ is shorter when the optical
pumping field is on compared to T$_1$ in the dark.

%
%
%
%

After optical pumping, the excitation laser was then deflected
with an acoustic optic modulator (AOM) for a time $\tau_{wait}$.
Next, a short pulse ($0.5-2\mu$s) resonant on the $A$ transition
probed the sample and light was detected from the $A^*$
transition.  We were able to perform the experiment on several
different D$^0$-D$^0$X transitions at high fields ($>$5~T).
However at low field it was only possible to cleanly isolate the
lowest energy $A$ transition and all data presented are from this
transition.

\begin{figure} [t]
{\includegraphics[height = 60 mm,
keepaspectratio]{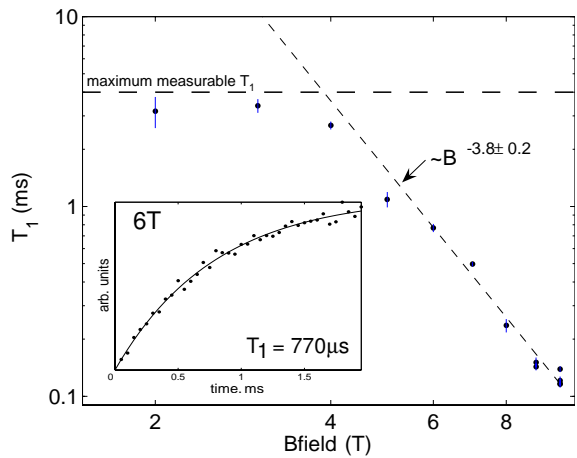}}\caption{Log-log plot of the bound
electron spin lifetime T$_1$ dependence on B-field.  A strong
power law dependence on B-field field is observed for B $>$ 5~T.
Recovery times measured at B $\leq$ 4~T are limited by the
experimental apparatus and are only a lower bound on the electron
T$_1$. Inset: Example exponential fit of the pump-probe data at
6~T.} \label{T1}
\end{figure}

The magnetic field dependence of the electron spin relaxation is
shown in Fig.~\ref{T1}.  At fields less than 4~T there appears to
be a levelling off of T$_1$ at several ms.  This is related to the
finite extinction ratio of the AOM ($\approx 1000$) and a small
leakage field even when the beam is deflected away from the
sample. Thus, for fields less than 4~T our data only gives a lower
bound on T$_1$. As the magnetic field is increased, T$_1$
decreases rapidly exhibiting a strong power-law dependence. The
T$_1$ dependence on B-field observed is very different from
previously reported T$_1$ measurements in $n$-GaAs by Colton et
al.~\cite{Colton04b} Previously T$_1$ was measured in higher doped
samples ($n = 3\times10^{15} \textrm{cm}^{-3}$) by time-resolved
polarization photoluminescence measurements on free excitons.
T$_1$ was shown to increase with magnetic field reaching a maximum
of 1.4~$\mu$s at the maximum attainable field of 5~T. This
different dependence may be due to the larger doping density
resulting in a greater interaction between neighboring donor-bound
electron spins.  A further study on the effect of doping density
on T$_1$ is necessary to understand the discrepancy between the
two experimental results.

%
%
%
%

Theoretical calculations of the B-field dependence of T$_1$ due to
the modulation of electron spin-orbit coupling by phonons predict
at B$^{-4}$ dependence of T$_1$ for neutral
donors~\cite{Pines57a}.  A fit to data for $B > 5$~T
(Fig.~\ref{T1}) shows a strong power law dependence with an
exponent of $m = -3.8 \pm 0.2$.  This indicates that the
one-phonon spin-orbit relaxation process is the dominant process
for GaAs neutral donors at high magnetic fields, g$\mu_B$B $>$
k$_B$T . In our system the temperature T is 1.5~K, the electron
g-factor is 0.41 \cite{Fu05a} and the low-field to high-field
transition occurs at B = 5.6~T.  Our results are similar to
previously reported quantum dot electron-spin relaxation times. In
GaInAs quantum dots, Kroutvar et al.~\cite{Kroutvar04a} found an
inverse power-law dependence of T$_1$ on magnetic field.  The
observed power law exponent, $\textrm{m}= -5$, was theoretically
predicted for disk-like quantum dots \cite{Khaetskii01a, Woods02a}
for single-phonon spin-orbit relaxation processes. The similarity
between the donor electron and quantum dot systems indicate that
in our sample electrons are extremely well-isolated and
non-interacting.

In conclusion, we have observed long, millisecond spin-flip times
of electrons bound to donors in bulk GaAs.  This direct pump-probe
measurement is possible due to the homogeneity of the D$^0$ and
D$^0X$ systems in high-purity GaAs.  Our result represents a
three-order of magnitude increase over previously measured D$^0$
electron T$_1$ times.  Both the long T$_1$'s observed and the
strong power law dependence of T$_1$ on magnetic field indicate
that donor-bound electrons are non-interacting in sufficiently
pure GaAs. Thus these systems are an attractive testing ground to
study electron spin dynamics for quantum information processes.

Financial support was provided by the MURI Center for photonic
quantum information systems (ARO/ARDA Program DAAD19-03-1-0199),
JST/SORST program (SPO30754) for the research of quantum
information systems for which light is used, and ``IT program''
MEXT, University of Tokyo. We would like to thank D.
Goldhaber-Gordon for the valuable discussions.

\bibliographystyle{apsrev}

\end{document}